\def\year{2018}\relax
\documentclass[letterpaper]{article} 
\usepackage{aaai18}  
\usepackage{times}  
\usepackage{helvet}  
\usepackage{courier}  
\usepackage{url}  
\usepackage{graphicx}  
\usepackage{xcolor}
\usepackage{amsfonts,amsthm,amsmath,amssymb}
\begin{document}
%
\title{Learning the Probability of Activation in the Presence of Latent Spreaders}
\author{Maggie Makar \\ CSAIL, MIT \\ Cambridge, MA\\ \url{mmakar@mit.edu}
\And John Guttag \\ CSAIL, MIT \\ Cambridge, MA \\ \url{guttag@mit.edu}
\And Jenna Wiens \\ CSE, University of Michigan \\ Ann Arbor, MI \\ \url{wiensj@umich.edu}
}

\maketitle
\begin{abstract}
When an infection spreads in a community, an individual's probability of becoming infected depends on both her susceptibility and exposure to the contagion through contact with others. 
While one often has knowledge regarding an individual's susceptibility, in many cases, whether or not an individual's contacts are contagious is unknown.
We study the problem of predicting if an individual will adopt a contagion in the presence of multiple modes of infection (exposure/susceptibility) and latent neighbor influence. 
We present a generative probabilistic model and a variational inference method to learn the parameters of our model. 
Through a series of experiments on synthetic data, we measure the ability of the proposed model to identify latent spreaders, and predict the risk of infection.  
Applied to a real dataset of 20,000 hospital patients, we demonstrate the utility of our model in predicting the onset of a healthcare associated infection using patient room-sharing and nurse-sharing networks. Our model outperforms existing benchmarks and provides actionable insights for the design and implementation of targeted interventions to curb the spread of infection. 
\end{abstract}

\section{Introduction and Background}
In the early 1900s, a New Yorker was identified as the first asymptomatic carrier of Typhoid. While she appeared healthy, ``Typhoid Mary,'' as she later became known, infected 51 individuals \cite{TyphoidMary}. 
A model estimating the probability of contracting Typhoid for a New Yorker in that era would have had to take into account three important factors: 1) the personal characteristics that would make her more or less susceptible to the infection ({\textit e.g}., genetic factors) 2) the network of individuals she has come into contact with, and 3) the contagious status of those individuals, regardless of their apparent healthy state. 
Mary's example highlights the difficulty of estimating the risk of being \textit{activated} ({\textit i.e}., getting infected) in the presence of both endogenous characteristics and exogenous factors that may be hidden. While we focus on infectious diseases, similar examples exist in other domains ({\textit e.g}., social networks).

We present a generative probabilistic model for modeling each individual's Probability of Activation in the presence of Latent Spreaders (PALS). PALS models the infection state of an individual as a random variable that depends upon both the individual's \textit{susceptibility}, which we assume is captured by observed individual-specific characteristics (e.g., age, medical history), and an \textit{exposure} state that cannot be directly observed but can be inferred based on one's network of contacts.


In the information diffusion literature, previous approaches that tackle tasks with missing data do not explicitly take into account susceptibility factors that make an individual more/less likely to get infected and do not model the spreader states based on neighbor characteristics \cite{hiddenhazard,btp,incompleteobs17,reconst}.      
Xing et al. \shortcite{CarinCollege} model susceptibility and infection as latent states but assume that influence can be exerted by infected nodes only. They conclude that their model could not identify ``asymptomatic shedders''. Fan et al. \shortcite{katherineHeller} model the overall exposure to the contagion (rather than the individual neighbor influence state) as a latent variable. They do not attempt to learn the neighbor-specific characteristics that make her a spreader. In economics, the workhorse of quantitative research pertaining to social interactions is the linear-in-means model. This model captures the notion that an individual's behavior depends on the average behavior and/or characteristics of members of her group. The linear-in-means model, which assumes that the spreader status of the neighbors is observed, is a special case of our proposed model. There have been several extensions to the classical linear-in-means model. Toulis and Kao \shortcite{lim_causl_net} consider situations where the influence is observed but the network structure is uncertain to derive a causal peer influence.
In contrast to previous approaches, we assume a fully observed network structure and individual characteristics, but make no assumptions about the neighbors' spreader states (latent or observed) and the mode of infection (through exposure or susceptibility). Our main contributions are:

\begin{enumerate}
\item \textbf{Improved risk estimation}: In situations where exposure plays a role, PALS outperforms the baseline model. Additionally, PALS is robust to different modes of infection. It accurately estimates the risk of infection if the infection is acquired through exposure only, susceptibility only ({\textit e.g}., non-communicable diseases), or a combination of both. 
\item \textbf{Identification of latent spreaders}: PALS accurately identifies influential individuals based on their characteristics, even in situations where this influence is not observed or recorded. We show that PALS applies to situations where the spreader states are partially observed.  
\item \textbf{Interpretability and actionability}: PALS learns weights that parameterize a mapping from individual characteristics and the exposure state to an infection state. By analyzing these weights, we can study factors that contribute to the infection and design interventions that can curb its spread. In addition, spreader states are modeled as functions of one's characteristics. This could prove useful in a clinical setting---by understanding which patients are most likely to spread the disease, physicians can preemptively choose whom to isolate.
\end{enumerate}
In a series of simulations, we show that PALS outperforms benchmarks and estimates an accurate probability of infection regardless of the infection mode. In situations where spreader states are partially observed, even greater gains in accuracy can be achieved. Applied to a real dataset of admissions at a large hospital, we demonstrate the practical utility of our model for 1) identifying potential asymptomatic carriers of a hospital associated infection, and 2) identifying which patients are likely to become infected.

\section{Generative model}
\begin{figure}[t]\label{plate_diag}
\centering
\includegraphics[scale=0.8]{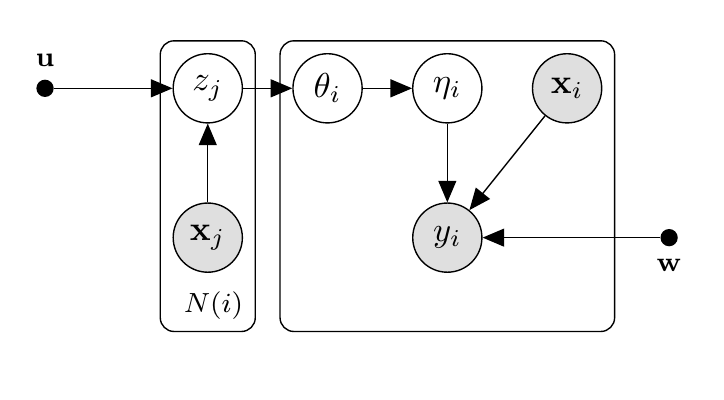}
\caption{Plate diagram for the proposed generative PALS model for an individual $i$. Infection is directly influenced by patient's characteristics and her neighbors' spreader states.}
\end{figure}

We model a population as a network in which nodes represent individuals and edges represent connections ({\textit e.g}., shared rooms or caregivers). 
We present a generative probabilistic model to estimate the Probability of Activation in the presence of Latent Spreaders (PALS).
Each neighbor's latent spreader state is modeled as a Bernoulli random variable, $z_j$, which takes on a value of 1 if she is spreading the infection and 0 otherwise. The parameter of the Bernoulli draw is computed as a function of the neighbor's vector of characteristics, $\mathbf{x}_j$, and a set of weights $\mathbf{u}$. For now, we assume that all spreaders are latent, and address situations where spreader states are partially observable in the next section. 
The probability that an individual $i$ has been exposed to the contagion, $\theta_i$, is encoded as a Beta variable parameterized by the number of contagious and non-contagious neighbors.\footnote{The parameters of the Beta distribution are shifted by one to ensure numerical stability.  One might incorporate some other informative prior over the number of spreaders/non-spreaders.}
This probability is then used as a parameter for a Bernoulli draw which defines the individual's exposure state, $\eta_i$. 
Finally, individual $i$'s infection state $y_i$, whether or not $i$ is infected, is drawn from a Bernoulli distribution. This distribution is parameterized by a probability computed as a function of $i$'s inherent susceptibility $\mathbf{x}_i$, $i$'s exposure state $\eta_i$ and a vector of weights $\mathbf{w}$.  
Putting the whole generative process together, for a single individual $i$:

\begin{enumerate}
\item For each neighbor $j \in n(i)$\\
Draw the spreader state: $z_j \sim$ Bernoulli($\sigma(\mathbf{u}^T \mathbf{x}_j))$ 
\item Draw the probability of exposure: \\ $\theta_i | \mathbf{z}_{i} \sim \text{Beta}( 1+ \sum_{n(i)} z_j, 1+ \sum_{n(i)} 1- z_j)$
\item Draw the exposure state: $\eta_i | \theta_i \sim \text{Bernoulli}(\theta_i)$
\item Draw the infection state: $y_i | \mathbf{x}_i, \eta_i \sim \text{Bernoulli}(\sigma(\mathbf{w}^T \mathbf{x}'_i))$
\end{enumerate}
$n(i)$ denotes the set of $i$'s neighbors, $\sigma(.)$ denotes the sigmoid function, $\mathbf{z}_{i}$ is a vector of the contagious states of $i$'s neighbors and $\mathbf{x}'_i$=$ \langle \mathbf{x}_i,\eta_i \rangle$, {\textit i.e}., the concatenation of an individual's characteristics and her exposure state. 
The joint probability distribution is expressed as:
\begin{align}
& \nonumber p(\mathbf{z}_i | \mathbf{u},  X_{n(i)}) p(\theta_i |\mathbf{z}_i) p(\eta_i | \theta_i)  p(y_i | \mathbf{x}_i, \eta_i, \mathbf{w}),
\end{align}
where $X_{n(i)}$ denotes the matrix of characteristics of all $i$'s neighbors (figure 1). In general, we use lowercase to denote scalars, boldface to denote vectors, and uppercase to denote matrices. 
This generative model captures several key aspects of the spread of infections. 
First, by modeling each individual spreader state, rather than the aggregate count of spreading neighbors,
we can make conclusions about each individual. This representation is both realistic and useful. It allows us to identify the culprits and target interventions. 
Second, by aggregating over each of the neighbor states and including exposure as a variable along with the susceptibility factors, we can directly infer the relative importance of exposure and susceptibility, reflected in the weight vector $\mathbf{w}$. Finally, by parameterizing the two Bernoulli draws in steps 1 and 4 as functions of the neighbor and individual characteristics respectively, our model gains the advantage of interpretability: both spreader and infection states are traceable to observed characteristics. By analyzing the model weights $\mathbf{u}$ and $\mathbf{w}$, we can understand how different factors affect both states.  

\section{Inference}
In this section, we give the details of the inference procedure used to learn the parameters of the model.  
Typically, one would use Expectation Maximization (EM) to find the values of these parameters. However, the E-step would require finding the posterior distribution over the latent variables, which requires evaluating: 
\begin{align}
& \nonumber \frac{p(\mathbf{z}_i | \mathbf{u}, X_{n(i)}) p(\theta_i | \mathbf{z}_i) p(\eta_i | \theta_i) p(y_i | \mathbf{x}_i, \eta_i, \mathbf{w})} 
{\int_\theta \sum_{\mathbf{z}} \sum_{\eta}p(\mathbf{z}_i | \mathbf{u}, X_{n(i)}) p(\theta_i | \mathbf{z}_i) p(\eta_i | \theta_i) p(y_i | \mathbf{x}_i, \eta_i, \mathbf{w})}.
\nonumber
\end{align}
The denominator is computationally intractable since it requires the evaluations of $2^{|n(i)| + 2}$ terms, so we resort to approximate inference.  For speed considerations, we chose variational inference \cite{jakola}. Specifically, we learn a set of parameters that parameterize a distribution $q$ over the latent variables such that it closely approximates the true posterior. We use mean-field variational inference, {\textit i.e.}, we restrict $\mathcal{Q}$ to the set of distributions that fully factorize:
{\fontsize{10.0}{11} \selectfont \begin{align}
q(\mathbf{z}_i, \theta_i, \eta_i) = \prod_{j \in n(i)} q(z_j | \phi_j) q(\theta_i | \gamma_i) q(\eta_i | \pi_i), 
\end{align}}
\noindent where each $q(z_j | \phi_j)$ and $q(\eta_i | \pi_i)$ are Bernoulli distributions, and $q(\theta_i | \gamma_i)$ is a Beta distribution. For purposes of easier notation, we express the Bernoulli distributions using two parameters---each $\phi_i$ is a vector of two parameters signifying the probability of being a spreader, $\phi_{i, 1}$, and the probability of being a non-spreader, $\phi_{i, 2} = 1- \phi_{i, 1}$. Similarly, $\pi_{i, 1}$ is the probability of being exposed, and $\pi_{i,2}$ is the probability of not being exposed. Let $\mathcal{D} = \{ \mathbf{x}_i, X_{n(i)}, y_i \}$ and $\Lambda = \{ \mathbf{z}_i, \theta_i, \eta_i \}$. We can now maximize the evidence lower bound (ELBO) which takes the expanded form:
{\fontsize{10.0}{11} \selectfont \begin{align} \label{var_obj}
& \nonumber \mathcal{L}(q) := \mathbb{E}_q[\log p(\mathcal{D}, \Lambda  | \mathbf{u}, \mathbf{w})] - \mathbb{E}_q[\log q(\Lambda)]  \\
\nonumber & = \mathbb{E}_q[\log p(\mathbf{z}_i| \mathbf{u}, X_{n(i)})]  + \mathbb{E}_q[\log p(\theta_i |\mathbf{z}_i)] \\
\nonumber  &  + \mathbb{E}_q[\log p(\eta_i | \theta_i)] 
 +\mathbb{E}_q[\log p(y_i | \mathbf{x}_i, \eta_i, \mathbf{w} )] \\
\nonumber  & -  \sum_{j \in n(i)}\mathbb{E}_q[\log q(\mathbf{z}_j | \phi_j)] - \mathbb{E}_q[\log q(\theta_i | \gamma_i)] \\
\nonumber & - \mathbb{E}_q[\log q(\eta_i | \pi_i)],
\end{align}}
\noindent where $\mathcal{L}$ is the ELBO, and $ \mathbb{E}_q$ is the expectation with respect to the variational distribution $q$. Two terms in the ELBO, $ \mathbb{E}_q[\log p(\theta_i |\mathbf{z}_i)] $ and $\mathbb{E}_q[\log p(y_i | \mathbf{x}_i, \eta_i, \mathbf{w} )]$, do not have a closed-form expression. In the next section we show how to evaluate them. The derivation of the other terms is presented in Appendix A.

\subsection{Evaluating the ELBO}
\textbf{Evaluating $\mathbb{E}_q[\log p(\theta_i |\mathbf{z}_i)]$}. Two of the terms that arise when expanding $\mathbb{E}_q[\log p(\theta_i |\mathbf{z}_i)]$ do not have a closed form solution. Specifically, we need to compute the terms  $-\mathbb{E}_q[\log \Gamma(\sum_j z_j)]$ and $-\mathbb{E}_q[\log \Gamma(\sum_j 1- z_j)]$. These terms cannot be evaluated analytically since they involve taking the expectation of the log of a nonlinear transformation of the summation of the latent variables. To untangle the sums over $z_j$, we utilize the following identity of the gamma function:
{\fontsize{10.0}{11} \selectfont \begin{align}
\log \Gamma(a + 1) = \log \Gamma(a) + \log (a). 
\end{align}}
\noindent Note that when writing the likelihood as a function of a single variable $z_j$, the two problematic terms that we need to compute the expectations over are expanded as follows:
{\fontsize{9.0}{10} \selectfont \begin{align}\label{seperating_sum}
\nonumber & - \log \Gamma( \sum_j z_j) - \log \Gamma( \sum_j 1- z_j) = \\
& - \log \Gamma(\sum_{k \not = j} z_k + z_j) -\log \Gamma(\sum_{k \not = j} 1- z_k + (1- z_j)). 
\end{align}}
\noindent We will take the expectation over Equation~\ref{seperating_sum} in two steps: first according to $q(z_j)$ then according to $q(\mathbf{z}_{\setminus j})$. First, since $z_j$  can only take on one of two values (1 or 0), taking the expectation over $z_j$, gives us the following expression:
{\fontsize{8}{9} \selectfont \begin{align}\label{kick_phi_out}
\nonumber & \mathbb{E}_{q(z_j)} \big[ - \log \Gamma( \sum_{k \not = j} z_k + z_j) -\log \Gamma(\sum_{k \not = j} (1- z_k) + (1- z_j)) \big] \\
\nonumber & = - \phi_{j, 1} \big[ \log \Gamma( \sum_{k \not = j} z_k) + \log(\sum_{k \not = j} z_k) \big]  - \phi_{j, 2}  \big[ \log \Gamma( \sum_{k \not = j} z_k) \big]  \\
 \nonumber &  - \phi_{j, 1} \big[ \log \Gamma( \sum_{k \not = j} 1- z_k) \big]   \\ & - \phi_{j, 2}  \big[ \log \Gamma(\sum_{k \not = j} 1- z_k) + \log( \sum_{k \not = j} 1- z_k) \big]. 
\end{align}}

\noindent Next, we take the expectation of this expression according to $q(\mathbf{z}_{\setminus j})$. This would once again entail taking the expectation over the log transformation of a sum of the variables and the log transformation of a nonlinear transformation of the sum of the variables. Here, following \cite{rtm,BraunMc} we use a first order Taylor approximation of equation~\ref{kick_phi_out} at $\mathbb{E}_{q(\mathbf{z}_{\setminus j})}[\sum_{k \not = j} z_k] = \sum_{k \not = j} \phi_{k} $. \\
\\ \textbf{Evaluating $ \mathbb{E}_q[\log p(y_i | \mathbf{x}_i, \eta_i, \mathbf{w} )] $}. The other expectation that is difficult to compute is
{\fontsize{9.0}{10} \selectfont \begin{align}\label{b_update2}
 & \nonumber \mathbb{E}_q[\log p(y_i | \mathbf{x}_i, \eta_i, \mathbf{w} )]  =  y_i   \mathbb{E}_q[\log \sigma(\mathbf{w}_{\setminus e}^T \mathbf{x}_i + w_{e} \eta_i) ]  \\
 & +  (1-y_i)  \mathbb{E}_q[\log (1- \sigma(\mathbf{w}_{\setminus e}^T \mathbf{x}_i + w_{e}  \eta_i) )], 
\end{align}}
\noindent where $\mathbf{w}_{\setminus e}$ is a vector of weights excluding the exposure weight and $w_{e}$ is the exposure weight. Equation \ref{b_update2} once again, requires computing the expectation of a log transformation of a nonlinear combination of the variables. Because $\eta_i$ can only take on two values: 1 and 0, we can compute this term as follows:
 {\fontsize{9.0}{10} \selectfont \begin{align}
& \nonumber \mathbb{E}_q[\log p(y_i | \mathbf{x}_i, \eta_i, \mathbf{w} )]  =  y_i [\pi_{i, 1}  \log \sigma(\mathbf{w}_{\setminus e}^T \mathbf{x}_i + w_{e})   \\
\nonumber & + \pi_{i, 2}  \log \sigma(\mathbf{w}_{\setminus e}^T \mathbf{x}_i) ]  +  (1-y_i)  [\pi_{i, 1}  \log (1- \sigma(\mathbf{w}_{\setminus e}^T \mathbf{x}_i + w_{e})) \\
\nonumber &   +  \pi_{i, 2} \log (1- \sigma(\mathbf{w}_{\setminus e}^T \mathbf{x}_i) )].
 \end{align}}
\noindent This term simply computes the probability of infection in case the individual is exposed and the probability of infection if she is not exposed then takes the weighted average of the two using the probability of exposure as the weight. 

Expanding equation~\ref{var_obj} and then maximizing, we get the updates of the variational distribution parameters and the global updates. \\
 \subsection{Variational E-step}
 In this step, we get the updates with respect to the free variational parameters that control our variational distributions.
\\ \textbf{Update with respect to $q(z_j)$}: 
{\fontsize{9.0}{10} \selectfont \begin{align} \label{phi_update}
&\nonumber \phi_{j, 1} \propto  \sigma(\mathbf{u}^T\mathbf{x}_j)  \exp ( \psi(\gamma_{j, 1}) ) ( 1+ \sum_{k \not = j } \phi_{1, k} )^{-1} \\
& \phi_{j, 2} \propto  (1- \sigma(\mathbf{u}^T\mathbf{x}_j))  \exp ( \psi(\gamma_{j, 2}) ) (1 + \sum_{k \not = j } \phi_{2, k} )^{-1}, 
\end{align}}
where $\psi$ is the digamma function, the first derivative of the log gamma function. This update has an intuitive explanation: the posterior probability of $j$'s spreader state is proportional to her probability of being a spreader computed according to her own characteristics, and the exposure state of the individual she is connected to but inversely proportional to the individual's other neighbors' states. This means that if individual $i$ is very likely exposed to the contagion, but her $\{ k \in n(i): k\not = j\}$ neighbors are very likely spreaders, the two last terms cancel out and the posterior spreader state becomes equal to the probability of being a spreader depending on $j$'s characteristics. In other words, the $\{ k \in n(i): k\not = j\}$ neighbors are sufficient to account for the individual's exposure and no ``blame'' is assigned to neighbor $j$. \\
\\ \textbf{Update with respect to $q(\theta)$}:
{\fontsize{9.0}{10} \selectfont \begin{align} 
\gamma_{i, s}  = \sum_{j \in {n(i)}}\phi_{j, s}  + \pi_{i, s}  + 1 && \text{for $s$ = 1, 2}.  
\end{align} }
The gamma update, which is interpreted as the variational parameter that controls the probability of exposure, depends on the individual's neighbors' spreader states and her exposure state. \\
\\\textbf{Update with respect to $q(\eta)$}:
{\fontsize{8pt}{9pt}
\selectfont 
\begin{align}
\nonumber \pi_{i, 1} &  \propto \sigma(\mathbf{w}_{\setminus e}^T \mathbf{x}_i + w_{e})^{y_i}  ( 1- \sigma(\mathbf{w}_{\setminus e}^T \mathbf{x}_i + w_{e}))^{1- y_i}   \exp(  \psi(\gamma_{i, 1})  ) \\
 \nonumber  \pi_{i, 2} & \propto \sigma(\mathbf{w}_{\setminus e}^T \mathbf{x}_i )^{y_i}  ( 1- \sigma(\mathbf{w}_{\setminus e}^T \mathbf{x}_i))^{1- y_i}   \exp (  \psi(\gamma_{i, 2})  ) . 
\end{align}}
The final exposure state also has an intuitive explanation: the $\pi_{i,1}$ update is proportional to the likelihood of infection given exposure and the probability of exposure, while the $\pi_{i,2}$ update is proportional to the likelihood of infection given non-exposure and the probability of non-exposure.

\subsection{Variational M-step}
This step gives us the updates with respect to the parameters $\mathbf{u}$ and $\mathbf{w}$. For a network with $N$ nodes, we have to find $\mathbf{u}$ and $\mathbf{w}$ that maximize the following two expressions:  
{\fontsize{9.0}{10} \selectfont
\begin{align}\label{u_update}
\mathcal{L}_\mathbf{u} &= \sum_{i=1}^{N} \sum_{j \in {n(i)}} \phi_{i, j, 1}\log \sigma(\mathbf{u}^T\mathbf{x}_j) \\
\nonumber & +  \phi_{i, j, 2} \log \big(1- \sigma(\mathbf{u}^T\mathbf{x}_j) \big),
\end{align}}
and 
{\fontsize{9.0}{10}  \begin{align}\label{w_update}
\nonumber \mathcal{L}_\mathbf{w} &  =  \sum_{i=1}^{N} y_i \pi_{i, 1} \log \sigma(\mathbf{w}_{\setminus e}^T \mathbf{x}_i + w_{e})  + y_i \pi_{i, 2} \log \sigma(\mathbf{w}_{\setminus e}^T \mathbf{x}_i) \\
\nonumber & + (1-y_i) \pi_{i, 1} \log(1 -  \sigma(\mathbf{w}_{\setminus e}^T \mathbf{x}_i +  w_{e})) \\
 & + (1-y_i) \pi_{i, 2}  \log (1 - \sigma(\mathbf{w}_{\setminus e}^T \mathbf{x}_i)). 
\end{align}}
These two expressions are similar to the traditional cross entropy of a logistic regression. However, in Equation~\ref{u_update} we have a probability instead of having a binary outcome. In Equation~\ref{w_update} the label is binary, but we take the weighted average over the probability of infection given exposure and the probability of infection given non-exposure weighted by the probability of exposure. 
These terms do not have a closed form solution, but they can be solved using a numerical method such as BFGS. 
Additionally, in situations where $\mathbf{x}$ is high dimensional, regularization can be incorporated, which we have found useful in practice. 

\subsection{Predictions}
We can use our model to make predictions about the spreader and/or infection states of out-of-sample data. Since the spreader state of an individual depends on only $\mathbf{u}$ and that individual's characteristics, computing a prediction for the spreader state is straightforward. 
To make predictions about the infection states, we need to compute the expected value of exposure. This requires a variational inference step. This step is identical to the one outlined previously with the exception that the terms that depend on the outcome $y$ are removed. 

We can extend PALS to situations where some spreader states are observed. In such a setting, we do not need to find the posterior over $q(z_j)$ for any $j$s where the spreader state is known during inference and prediction. 
\section{Synthetic Experiments and Results}
Through a series of experiments on synthetic data, we evaluate the performance of our method under a variety of different conditions. Synthetic data allow us to evaluate how well the method does at identifying latent spreaders, which, by definition, we would not have ground truth for in real data. \\ 
\subsection{Experimental setup}
\textbf{Generating the network}. For all experiments in this section, we generate a network of 500 individuals for training and a held-out network of 500 individuals for testing. 
We use a stochastic block model (SBM) to simulate the network and set the probability that two individuals within and across a sub-community form an edge is 0.5, and 0.01 respectively. We chose a SBM because it allows us to create sub-communities and flexibly define the probability of forming edges within and across sub-communities.
For simplicity, instead of drawing $\theta$ as a stochastic function of $\mathbf{z}$, we compute it as the mean of the spreader states and set $\eta = 1$ if $\theta \geq 0.5$.
We generate a vector of characteristics of 20 binary variables for each individual. 
Further details about the data generation process and additional experiments are available in Appendix B. \\
\\
\textbf{Evaluation metrics and benchmarks}. We report performance on the held-out test set, averaged over 30 runs. We use the Area Under the receiver operating Curve (AUC) to measure the performance of PALS in predicting the spreader state and the infection state. We measure two AUC's: $\mathbf{y}$-PALS, which measures the accuracy in predicting the infection state and $\mathbf{z}$-PALS, which measures the accuracy in predicting the spreader state.
We compare our results to the following benchmarks: 
\begin{itemize}
\item NoNet: A logistic regression model that ignores the network, estimating an individual's risk of infection $y_i$ using only individual characteristics $\mathbf{x}_i$. When exposure is irrelevant, PALS should perform identically to NoNet.
\item $\eta_\mathcal{O}$: A logistic regression model that has oracle access to the neighbor spreader states, and hence has access to the ground truth of whether or not an individual is exposed in addition to the individual's characteristics. This model presents a best-case scenario which is almost never available in a real setting. 
\item $\mathbf{z}_\mathcal{O}$: A logistic regression model that has oracle access to the ground truth label of the spreader states, $\mathbf{z}$, during training time. At test time, it predicts the spreader states based on individual characteristics and learned weights. This provides a benchmark for the accuracy in predicting the spreader state.
\end{itemize}

\begin{figure}
\centering
\includegraphics[scale=0.34]{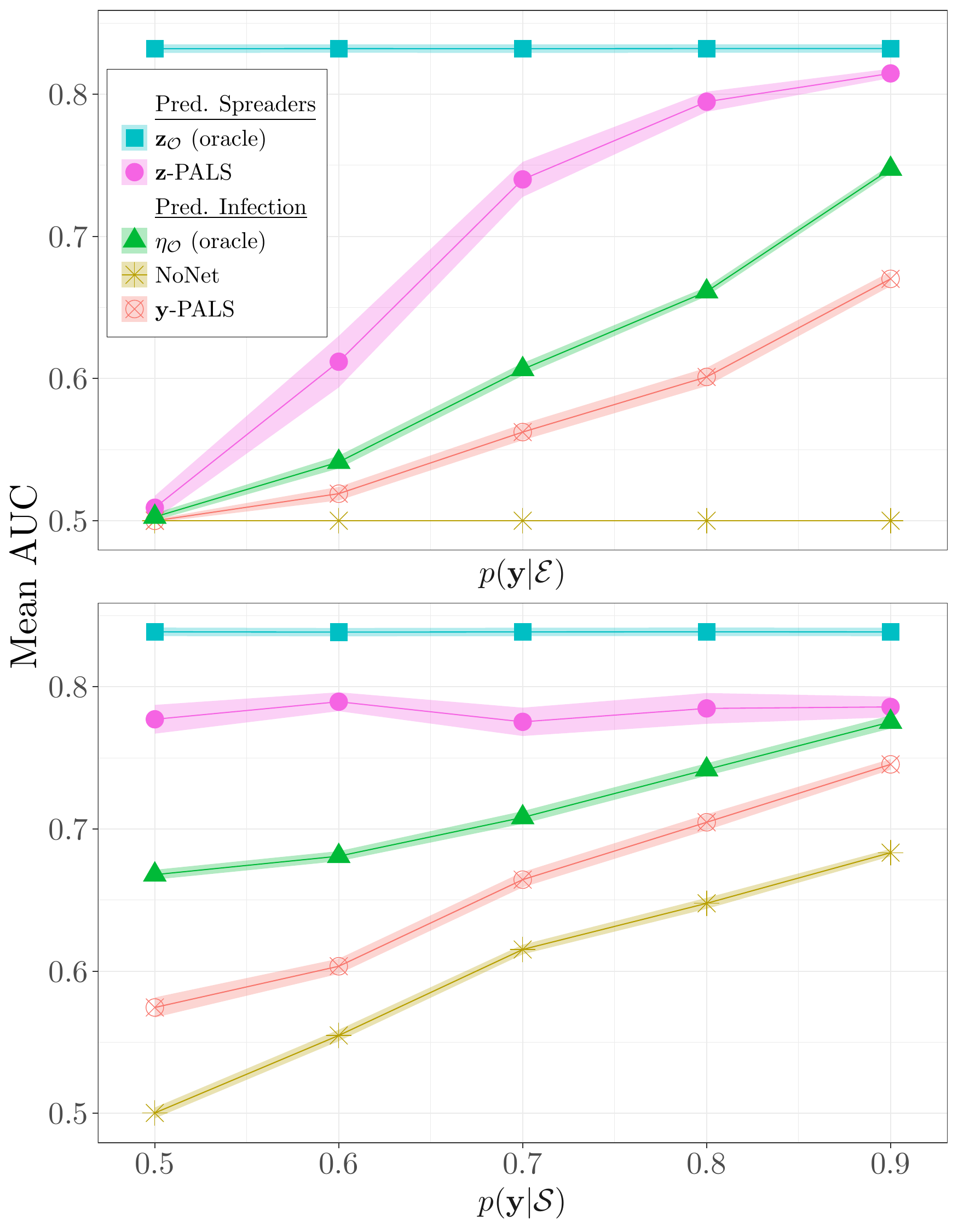} 
\caption{(Top) PALS outperforms NoNet when exposure is important. (Bottom) PALS outperforms NoNet as susceptibility changes.}
\end{figure}
\subsection{Experiments and Results}
\subsubsection{Experiment 1: varying the probability of infection conditional on exposure.} \textit{How does varying the importance of exposure in determining infection, keeping all else constant, affect the accuracy of PALS?}
To focus on the effect of exposure, we keep the probability of infection conditional on susceptibility, $p(\mathbf{y}| \mathcal{S})$, constant and equal to 0.5. By doing so, we ``turn off" the effect of susceptibility. We vary the probability of infection conditional on exposure, $p(\mathbf{y}| \mathcal{E})$, from 0.5 to 0.9 and measure changes in accuracy; the graph is symmetric about 0.5.

The top panel in Figure 2 shows the results from Experiment~1. At $p(\mathbf{y}| \mathcal{E}) = 0.5$, the infection state is akin to a coin flip, since $p(\mathbf{y}| \mathcal{S})$ is also 0.5 by design. As expected, we see that all three models perform equally poorly when predicting the infection state (AUC = 0.5). However, as $p(\mathbf{y}| \mathcal{E})$ increases, $\eta_\mathcal{O}$, which has access to the ground truth of the spreader states (and hence the exposure states) makes more accurate predictions about the infection state. On the other hand, NoNet, which ignores exposure continues making predictions only as good as random regardless of the value of  $p(\mathbf{y}| \mathcal{E})$. Importantly, we see that $\mathbf{y}$-PALS improves as $p(\mathbf{y}| \mathcal{E})$ increases. We also see that $\mathbf{z}$-PALS improves as  $p(\mathbf{y}| \mathcal{E})$ increases, even performing comparably to $\mathbf{z}_\mathcal{O}$ at high values of $p(\mathbf{y}| \mathcal{E})$. 

\subsubsection{Experiment 2: varying the probability of infection conditional on susceptibility.} \textit{How does varying the level of importance of susceptibility in determining infection, keeping all else constant, affect the accuracy of PALS?}
Similar to the previous experiment, we keep $p(\mathbf{y}| \mathcal{E})$ constant and vary  $p(\mathbf{y}| \mathcal{S})$. Unlike experiment 1, we do not set  $p(\mathbf{y}| \mathcal{E}) = 0.5$ since there is no utility to using PALS in a setup where exposure does not affect the outcome. Instead, we set $p(\mathbf{y}| \mathcal{E}) = 0.8$. Furthermore, to elucidate the utility of PALS, we impose an upper bound on the percent of individuals who can get the infection through susceptibility to be 50\%. This in turn means that 50\% of the population can only get infected through exposure. We vary $p(\mathbf{y}| \mathcal{S})$ from 0.5 to 0.9 only for the 50\% of the population that is susceptible.  

The bottom panel in Figure 2 shows the results from Experiment~2. When $p(\mathbf{y}| \mathcal{S}) = 0.5$, infection is independent of susceptibility and only exposed individuals are infected. We clearly see that $\mathbf{y}$-PALS is better than NoNet. As $p(\mathbf{y}| \mathcal{S})$ increases, the performance of all three models improves but NoNet is consistently lower than $\mathbf{y}$-PALS and $\eta_\mathcal{O}$. This is because, by design, only a maximum of 50\% of the sample can be detected by NoNet. Importantly, we notice that $\mathbf{z}$-PALS is unaffected by changes in $p(\mathbf{y}| \mathcal{S})$. 

Results from the first two experiments suggest that the accuracy in predicting the spreader states is unaffected by the degree to which susceptibility decides infection but, as expected, is affected by how important exposure is in deciding the infection state. This dependence is desirable: if exposure is not important in determining the infection state, the entire concept of detecting spreaders is useless and our model appropriately defaults to one that ignores the exposure state.  

\subsubsection{Experiment 3: varying levels of spreader observability.} \textit{How does the accuracy of PALS change as we incorporate partially observed spreader states?}
This experiment highlights scenarios where PALS is most useful and explores whether or not the accuracy improves when some of the spreaders are known. 
This setup is similar to Experiment~1, with $p(\mathbf{y}| \mathcal{E}) := 0.8$ and a slightly harder spreader prediction task, where the spreaders are more dispersed among the community. This is to study the impact when some proportion of the spreaders is known. Here, we vary the proportion of known spreaders from 0 to 1. We present results when the proportion of spreaders is known at training time only ($\mathbf{y}$-PALS-T and $\mathbf{z}$-PALS-T) and at both training and testing time ($\mathbf{y}$-PALS-TT)\footnote{Since there is no utility to predicting the spreader state when it is known, we do not present $\mathbf{z}$-PALS-TT.}. We introduce an additional benchmark $\mathbf{\eta}_{\mathcal{O}(k)}$ for this experiment, where  the subscript ${\mathcal{O}(k)}$ denotes oracle access to the ground truth state of $k$\% of the spreaders. This is a logistic regression with oracle access to the ground truth spreader states of only $k\%$ of the population (rather than 100\% for $\eta_\mathcal{O}$). For the remaining $100-k\%$, the model assumes that they are non-spreaders, which mimics real world situations. For example, when a patient is asymptomatic, she is assumed to be healthy and not spreading the disease. 

Figure 3 shows the results from Experiment~3. The top and bottom panels show results from the infection and spreader prediction tasks respectively.
We see that as the proportion of known spreaders increases PALS makes more accurate predictions for both tasks.
Specifically, we see the greatest increase in accuracy when ground truth states are available at training and testing time. Beyond 30\% known spreaders, increasing the proportion of known spreaders at training time does not affect the accuracy of $\mathbf{y}$-PALS because it is already able to perfectly predict the spreader state. 
This shows that PALS can be easily extended to incorporate observed spreader labels which in turn leads to better performance.

Additional experiments studying how PALS performs under varying levels of sparsity of the graph and dispersion of the spreaders are presented in Appendix B. 

\begin{figure}[t!]\label{setup3fig}
\centering
\includegraphics[scale=0.34]{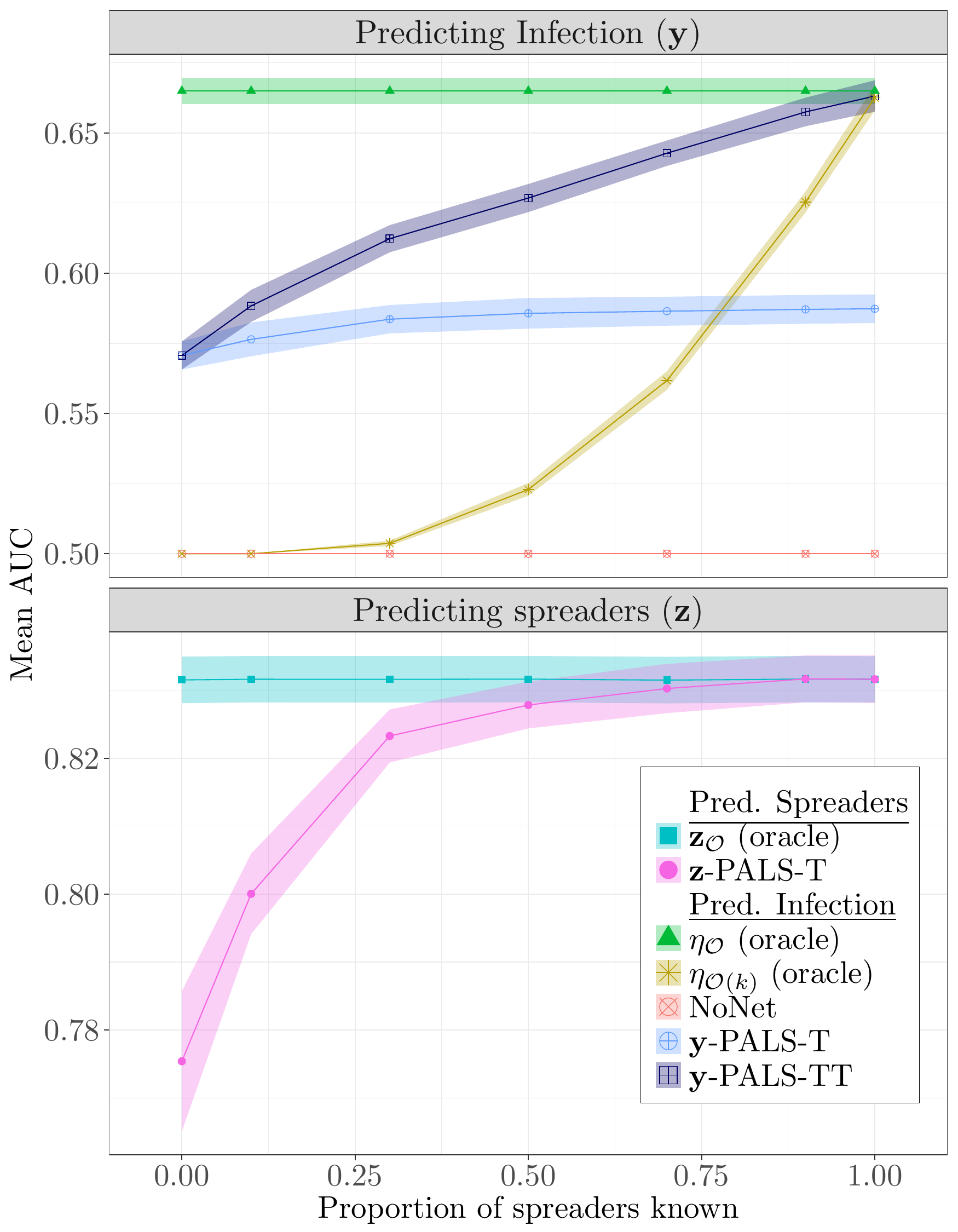}
\caption{Incorporating known spreader states leads to gains in accuracy for both infection and spreader prediction tasks.}
\end{figure}

\section{Real Data Experiments}
\textit{Clostridium difficile} (\textit{C. diff}) is responsible for over 300,000 healthcare-associated infections per year \cite{cdiff_nejm}. To acquire a \textit{C. diff} infection (CDI), a patient must be both susceptible ({\textit e.g.}, immunocompromised) and exposed to the disease ({\textit i.e.}, ingest \textit{C. diff} spores excreted through an infected patient's stool). A patient may become exposed to the disease either before, during, or after a hospital admission. Importantly, even though medical experts hypothesize that asymptomatic carriers of \textit{C. diff} contribute to the spread of the infection, the Center for Disease Control recommends only testing symptomatic patients, since there is no principled way of identifying potential asymptomatic carriers \cite{asymptomCdiff,cdcguide}. Existing methods for predicting the onset of CDI assume that only symptomatic carriers can spread the disease, and do not consider patient specific characteristics that make them likely spreaders \cite{wiens2014learning,wiens2016,dub}. PALS is therefore suitable for this task, since it would help identify asymptomatic carriers and compute a better estimate of exposure. 
We applied PALS to a dataset of patient admissions to a large urban hospital.  We consider the task of predicting who is most likely to be diagnosed with CDI during a hospitalization, while explicitly modeling asymptomatic carriers as latent spreaders. \\

\noindent \textbf{Main patients}. We consider inpatient hospitalizations that occurred from May 2012 to May 2014. After applying exclusion criteria, 20,147 admissions remained.  These criteria (outlined in detail in the Appendix C) exclude young patients and short-stays. We split the data into training and testing temporally, using hospitalizations from May 2012 to May 2013 as training data. Our outcome of interest is a binary label indicating whether or not the patient was diagnosed with CDI after the fifth day of their visit but before discharge. We chose the fifth day as a prediction date in order to focus on infections that are most likely hospital-associated rather than infections that may be attributable to contact outside the hospital. Since we are interested in prediction, we exclude patients who test positive for CDI prior to the fifth day of their visit. In our final population, there are 305 cases of CDI.

\noindent\textbf{Auxiliary patients}. We include patients who do not meet our inclusion criteria but who come into contact with our study population as auxiliary patients. 
We are only interested in predicting their spreader states, not their infection states. We consider them only as potential spreaders who come into contact with the main cohort. 

\noindent\textbf{Constructing the network}. We construct two networks consisting of patients who come into contact with the main patients. In the \textit{room} network, two patients are connected by an edge if they spend any time in the same room during the same day.  In the \textit{nurse} network, patients are connected if they have drugs administered by the same nurse on the same day. The nurse network is denser than the room network (median of 12 vs. 2 contacts per patient), because room changes are less frequent than nurse visits. Additional statistics about the population are included in Appendix C.

\noindent\textbf{Extracting patient characteristics}.  We extract a set of variables available upon admission such as demographic details, and medical history. We also extract data about the patient's current visit such as procedures, medications, laboratory tests and values and hospital location (units) occurring prior to the maximum extraction date.  All non-binary variables are binned into quintiles and made binary. 
A single patient can have multiple extraction dates corresponding to multiple views: 1) as the ``main'' patient for whom we are trying to make predictions, and 2) as a ``neighbor'' of another patient for whom we are trying to make predictions. In the first role, we extract information about the patient up until the fifth day of admission. In the second, we consider data about the patient only up until the date of contact with the main patient under consideration. 

Extracting the data in this way respects the causal ordering of events and does not leak information that would be unavailable at the time of prediction. Additionally, we exclude medications meant to treat CDI from the patients' information when they are in the first role, but include them when they are in the neighbor role\footnote{Even though all the patients in this cohort have not tested positive for CDI before the fifth day, occasionally physicians will start treating the patient before receiving the conclusive test results.}. We refer to the collection of main patient, room-sharing and nurse-sharing views as the \textit{main}, \textit{room}, and \textit{nurse} cohorts, respectively. \\

\noindent\textbf{Benchmarks and Models}\\
We compare PALS to models that take into account susceptibility, exposure and a combination of the two. To do so, we construct several proxies for exposure: 
 \begin{itemize}
\item Neighbor infection (NbrInf): this exposure measure assumes that there are no asymptomatic carriers, and computes the patients' exposure as the mean number of infected patients she has come into contact with. This is similar to the colonization pressure approach presented in \cite{wiens2014learning,wiens2016}. 
\item Neighbor infection rate (NbrInfRate): takes the contacts' infection \textit{rates} to be a proxy for exposure. If main patient A is in contact with B, we count the proportion of contacts that B has, excluding A, who have acquired CDI. We then take the average of all of patient A's contacts' rates as the exposure measure. 
\item Neighbor's probability of infection (NbrProbInf): takes the contacts' probabilities of infection as a proxy for exposure. We use the contacts cohorts' characteristics to predict whether or not they will get CDI. We then average over all the predicted probabilities of infection for all contacts of a given patient to estimate her exposure. 
\end{itemize}
For the exposure-only benchmarks, we use the exposure as a direct estimate of the main patients' probability of infection. For the exposure and susceptibility benchmarks, we add the exposure estimate to the main patients' characteristics and use an L1-regularized logistic regression to learn the parameters.
We run two versions of PALS considering each of the two networks. The first version, (NoObs), assumes that spreaders are completely unobserved; the second (PartObs), uses the CDI test values of the patients who got tested ($\approx$10\%) as observed spreader labels, and assumes the remaining are unobserved. \\

\noindent\textbf{Results} \\
Table 1 shows performance on the held-out test sample in terms of AUC for the task of predicting CDI. Results from the exposure-only and susceptibility-only models imply that, while exposure has a good predictive power, susceptibility-only variables tend to dominate in terms of performance. We find that the benchmarks that incorporate both susceptibility and exposure are not able to leverage the predictive power of exposure, as evidenced by the fact their performance is nearly identical to the susceptibility-only model. The greatest improvements over the susceptibility-only model are achieved by the PALS-PartObs (nurse) and the PALS-NoObs (room) respectively (0.705 and 0.704 compared to 0.7 for the best benchmark). 

In addition to the AUC, we also consider the TPR at FPR = 0.1. If a model is used to help decide when an intervention ({\textit e.g}., testing or isolating a patient) should be undertaken, a decision threshold must be chosen. The appropriate threshold depends upon the intervention. In this evaluation, we chose a cutoff corresponding to a false positive rate (FPR) of 0.1. This is appropriate because the cost of a typical intervention for CDI (\textit{e.g}., isolating the patient) is high. A threshold with a high FPR would lead to wasting valuable resources treating and testing patients who are not going to get infected.  We find that PALS-PartObs (room) has the highest TPR, which is statistically significantly higher than all benchmarks. Specifically, it is 2\% higher than the best benchmark, the model that combines the room contacts' infection state and the main patient's susceptibility characteristics.  
Note that 2\% of 300,000 (the number of hospital-associated CDI infections per year in the US) is 6,000 patients whose treatment might be changed. Furthermore, we find that our conclusions hold for other FPRs. 
At FPR=0.01, 0.05 and 0.5, PALS models achieve TPRs equal to 0.034 (95\% CI: 0.033, 0.035), 0.149 (95\% CI: 0.147, 0.151) and 0.78 (95\% CI: 0.777, 0.782) respectively. Benchmark TPRs are 0.023 (95\% CI: 0.02, 0.024), 0.14 (95\% CI: 0.138, 0.142), and 0.77 (95\% CI: 0.767, 0.772). In all cases, the PALS models outperform benchmarks. 

In addition to improved predictions, inspecting $\mathbf{u}$ gives us insight into the factors that lead to an elevated risk of being a spreader. Analyzing the $\mathbf{u}$ weights from our best performing model, PALS-PartObs (nurse), we find that receiving treatment for CDI is associated with the largest \textit{negative} weight, indicating that that patients being treated for CDI are not spreading the pathogen. This implies that the hospital's contact precautions are effective in curbing the spread of the infection from patients who have been clinically diagnosed with CDI. 
We also find that the highest positive $\mathbf{u}$ weights are those associated with broad-spectrum antibiotics (IV-Vancomycin, Ciprofloxacin, and Ceftriaxone) and treatment for diarrhea (Loperamide). Broad spectrum antibiotics are known to create an environment where \textit{C. diff} flourishes, and diarrhea increases the spread of \textit{C. diff}.
These results suggest that asymptomatic patients who are on antibiotics, have diarrhea, and are not being treated for CDI may be shedding \textit{C. diff} spores and putting their neighbors at a higher risk of CDI. 

\begin{table}
\centering
\scalebox{0.8}{
\begin{tabular}{lcc} 
 \hline
 & AUC & TPR at FPR = 0.1 \\
  \hline
\multicolumn{3}{c}{\textbf{Exposure-only (room)}}  \\
  \hline
\ \ \ \ NbrInf & 0.507 (0.506, 0.507) & 0.110 (0.110, 0.110)  \\
\ \ \ \ NbrInfRate & 0.503 (0.502, 0.503) & 0.107 (0.105, 0.108) \\
\ \ \ \  NbrProbInf & 0.585 (0.584, 0.587) & 0.153 (0.151, 0.155) \\
  \hline
\multicolumn{3}{c}{\textbf{Exposure-only (nurse)}}  \\
  \hline
\ \ \ \ NbrInf & 0.543 (0.542, 0.545) & 0.109 (0.108, 0.111) \\
\ \ \ \ NbrInfRate & 0.606 (0.605, 0.608) & 0.127 (0.125, 0.129) \\
\ \ \ \ NbrProbInf & 0.641 (0.639, 0.642) & 0.196 (0.194, 0.198) \\
  \hline
\multicolumn{3}{c}{\textbf{Susceptibility-only }}\\
  \hline
 & 0.698 (0.694, 0.703)  & 0.298 (0.296, 0.300) \\
   \hline
\multicolumn{3}{c}{\textbf{Exposure (room) + Susceptibility }} \\
  \hline
\ \ \ \ NbrInf & 0.697 (0.695, 0.698) & 0.292 (0.289, 0.294) \\
\ \ \ \ NbrInfRate & 0.694 (0.693, 0.695) & 0.263 (0.260, 0.265) \\
\ \ \ \ NbrProbInf & 0.697 (0.696, 0.698) & 0.278 (0.276, 0.280) \\
  \hline
\multicolumn{3}{c}{\textbf{Exposure (nurse) + Susceptibility}}  \\
  \hline
\ \ \ \ NbrInf & 0.694 (0.693, 0.696) & 0.264 (0.262, 0.266) \\
\ \ \ \ NbrInfRate & 0.700 (0.699, 0.702) & 0.300 (0.297, 0.302) \\
\ \ \ \ NbrProbInf & 0.695 (0.694, 0.696) & 0.278 (0.276, 0.281) \\
  \hline
\multicolumn{3}{c}{\textbf{PALS (room) } }\\
  \hline
\ \ \ \  NoObs & 0.704 (0.702, 0.705) & 0.299 (0.296, 0.302) \\
\ \ \ \  PartObs & 0.701 (0.700, 0.702) & {\bf 0.324 (0.322, 0.327)} \\
  \hline
\multicolumn{3}{c}{\textbf{PALS (nurse)}} \\
  \hline
\ \ \ \  NoObs & 0.700 (0.699, 0.702) & 0.298 (0.296, 0.301) \\
\ \ \ \ PartObs & {\bf 0.705 (0.703, 0.706)} & 0.310 (0.308, 0.313) \\
\hline
\end{tabular}
}
\caption{Predicting onset of CDI, test set performance. Benchmarks incorporating both susceptibility and exposure are not able to leverage the predictive power of exposure to the same extent as PALS.}
\centering
\end{table}
\section{Conclusion and Future Work }
We presented a novel and computationally efficient method, PALS, for inferring the latent influence of neighbors, estimating exposure and predicting risk of infection based on both exogenous exposure and inherent susceptibility. 
In a series of simulations, we demonstrated that PALS was strictly better than a baseline method that does not take into account exposure to the contagion. When the probability of infection was independent of exposure, PALS' performance was the same as the baseline method. But when the probability of infection could be affected by neighbors, PALS had better predictive performance than the baseline method. Additionally, we showed that PALS can accurately identify latent spreaders of infection. 

We also tested PALS on data from a large hospital, and used it to build a predictive model for \textit{C. diff} infections.
Our model outperformed all other benchmarks in predicting CDI.
Unlike existing work, our model sheds light on how the infection is spread within the hospital. This enables targeted interventions, designed to reduce the overall prevalence of the disease, increasing the actionability of the model. 

Though we present PALS in the context of infectious disease, our approach is widely applicable. For networks in which transmission dynamics are governed by latent spreaders and underlying node characteristics, PALS can accurately model the likelihood of adopting the contagion while shedding light on what makes a node more likely to transmit. 

\subsubsection{Acknowledgments}
This work was supported by the NSF award number IIS-1553146, the NIAID of the NIH (U01AI124255) and NIH award P50-0267666-0002. The views and conclusions in this document are those of the authors and should not be interpreted as necessarily representing the official policies, either expressed or implied, of the NSF nor the NIH.
The authors would like to thank Dr. Erica Shenoy, Dr. David Hooper, Erin Ryan, Lauren West, Christopher Fusco, Robert McCaffrey, Karina Bradford, Joseph Marchesani, Keith Jennings, Davis Blalock, Jen Gong and the anonymous reviewers for their help.

\fontsize{9.0pt}{10.0pt}
\selectfont
\bibliography{pals}
\bibliographystyle{aaai}
\end{document}